# Surface waves in defocusing thermal nonlinear optical media

Yaroslav V. Kartashov, Fangwei Ye, Victor A. Vysloukh,* and Lluis Torner

*ICFO-Institut de Ciencies Fotoniques, and Universitat Politecnica de Catalunya, Mediterranean Technology Park, 08860 Castelldefels (Barcelona), Spain*

We predict that the interface of materials with *defocusing thermal* nonlinearities support stable fundamental and higher-order surface waves when the opposite edges of the medium are maintained at different temperatures. Such surface waves exist due to the interplay between repulsion from the interface and the defocusing thermal nonlinearity that deflect light beams from the bulk of the medium toward its edges.

Nonlocality of the nonlinear response is typical for many materials and it arises when such nonlinearity mechanisms as diffusion of carriers, reorientation of molecules, or heat diffusion are involved [1,2]. Nonlocality drastically affects the interactions between solitons [3-8]. Among the materials possessing nonlocal response are thermal materials [9-14]. Light propagation in such media is affected by the geometry of the sample and by the temperature distribution at its boundaries. Most previous studies addressed materials with positive thermal coefficients; nevertheless, in practice many thermal materials exhibit negative coefficients (such situation is encountered, e.g., in dye solutions [15] and liquid crystals with thermal nonlinearities [16]). In this case, bright solitons can not form in bulk samples, thus radiation is deflected toward the boundaries. Localization of light at the boundaries may lead to the excitation of surface waves [17,18]. In nonlocal media surface waves have been studied only for the case of focusing [19-21]. Recently, two-dimensional surface soliton waves were observed experimentally in lead glass with focusing thermal nonlinearity [22]. In this Letter we predict that localized surface waves can exist at the interface of thermal media featuring a defocusing nonlinearity, when the sample width and the boundary temperatures substantially affect the entire refractive index distribution. We find that both fundamental and higher-order surface solitons are stable in such settings.

We consider the propagation of laser beam along the $\xi$ axis in the vicinity of the interface of a thermal medium occupying the region $0 \leq \eta \leq L$. The propagation of light is described by the following system of equations [9,10,14]:

$$i\frac{\partial q}{\partial \xi} = -\frac{1}{2}\frac{\partial^2 q}{\partial \eta^2} - qn, \quad \frac{\partial^2 n}{\partial \eta^2} = |q|^2, \text{ for } 0 \leq \eta \leq L$$
$$i\frac{\partial q}{\partial \xi} = -\frac{1}{2}\frac{\partial^2 q}{\partial \eta^2} - qn_{\mathrm{d}}, \text{ for } \eta > L \text{ and } \eta < 0 \tag{1}$$

Here $q = (k_0^2 x_0^4 \alpha |\beta|/\kappa n_0)^{1/2} A$ is the dimensionless light field amplitude; $n = k_0^2 x_0^2 \delta n / n_0$ is proportional to the nonlinear change $\delta n$ of the refractive index $n_0$; $\alpha, \beta, \kappa$ are the optical absorption coefficient, the thermo-optic coefficient $(\beta = dn/dT < 0)$, and the thermal conductivity coefficient, respectively; the transverse and longitudinal coordinates $\eta, \xi$ are scaled to the beam width $x_0$ and the diffraction length $k_0 x_0^2$, respectively; $n_{\mathrm{d}} < 0$ describes the difference of the unperturbed refractive indices of the thermal and the surrounding medium. Here we address the steady-state regime when the temperature distribution does not change with time. Such regime is achieved when the boundaries of the medium (located at $\eta = 0$ and $\eta = L$) are kept at fixed temperatures. Note that otherwise space-time dynamics may introduce important new effects (see, e.g., Ref. [9]).

On physical grounds, the light beam is slightly absorbed upon propagation in the thermal medium and acts as a heat source. Heat diffuses creating a non-uniform temperature distribution, where a refractive index variation is proportional to the temperature change in each point. When $\beta < 0$ heat diffusion results in a local increase of $n$ near the sample boundaries and radiation is deflected towards the boundaries, a phenomenon which may result in surface wave excitation. When the sample boundaries are kept at equal temperatures $(n|_{\eta=0} = n|_{\eta=L})$, such waves emerge in pairs at the opposite boundaries. However, keeping one of the sample boundaries at a lower temperature (which is physically equivalent to creating a local increase of refractive index in the vicinity of this boundary) facilitates light localization near this boundary. This is illustrated in Fig. 1, which shows the propagation of Gaussian beams in thermal media. Figure 1(a) corresponds to a beam launched at the center of the sample, and Fig. 1(b) corresponds to a beam launched closer to its left boundary, which is kept at a lower temperature than the right boundary. This results in a linear increase of refractive index toward the left boundary in the absence of light, that causes deflection of the input beams toward this boundary. The amplitude of the near-surface oscillations shown in the plots substantially decreases with decrease of initial separation between beam center and the surface.

We assume that $n|_{\eta=0} = n_{\rm b} > 0$, while $n|_{\eta=L} = 0$. Note that one can set $n|_{\eta=L} = 0$ because adding a constant background in the refractive index is equivalent to introducing a shift of the soliton propagation constant. The width of the thermal medium is set to $L = 40$ and $n_{\rm d} = -100$ is small enough to account for possible differences in refractive indices of linear and nonlinear materials. For $|n_{\rm d}| \gg 1$ surface waves almost do not penetrate into the linear medium. We assume that the temperature of the surrounding linear medium is constant at $\eta < 0$ and $\eta > L$ and that the temperature variation does not cause any modifications in the linear medium. The thermal contribution to refractive index $n$ is given by $n(\eta) = \int_0^L G(\eta,\lambda)|q(\lambda)|^2 d\lambda + n_{\rm b}(1 - \eta/L)$, where $G(\eta,\lambda) = \eta(\lambda - L)/L$, for $\eta \leq \lambda$, and $G(\eta,\lambda) = \lambda(\eta - L)/L$, for $\eta \geq \lambda$, is the response function of the thermal medium. For $q \equiv 0$ the refractive index $n$ increases linearly toward the left boundary (Fig. 2). When a light beam is launched in the vicinity of the interface it modifies the refractive index shape because $n$ decreases in the heated regions. Note that the refractive index profile is distorted in the entire sample and not only in the narrow region of the order of beam width. The higher the beam intensity the stronger the reduction of the refractive index, but even for high-intensity beams the refractive index remains locally enhanced in the region adjacent to the left boundary; hence a surface wave can form.

We searched for profiles of surface waves in the form $q = w(\eta)\exp(ib\xi)$, where $b$ is the propagation constant. The interface under study supports nodeless fundamental waves (Fig. 2(a)), dipole (Fig. 2(b)), triple-mode waves (Fig. 2(c)), and a variety of higher-order structures. For higher-order modes the wave pole farthest from the interface always features the highest amplitude. All such waves exist below an upper cutoff $b_{\rm co}$ on $b$. With decreasing $b$ the peak amplitude of surface waves increases and the distance between the intensity maximum and the interface decreases. For all types of solutions the energy flow $U = \int_{-\infty}^{\infty} |q|^2 d\eta$ is a monotonically decreasing function of $b$, so that $U \to 0$ when $b \to b_{\rm co}$ (Fig. 3(a)). The width of the low-amplitude surface waves does not diverge near cutoff but remains finite. This width is determined by the gradient of refractive index profile and decreases with increase of $n_b$ at fixed $L$. The cutoff monotonically increases with $n_b$, while $b_{co}$ is always smaller than $n_b$. In Fig. 3(b) we show the difference $n_b - b_{co}$ versus $n_b$ for fundamental and dipole surface waves. The cutoff for the fundamental surface waves always exceeds that for dipole and higher-order modes. Note that we did not find localized surface modes in analogous settings but with focusing thermal nonlinearities [13].

To elucidate the stability of surface waves we searched for perturbed solutions of Eq. (1) with the form $q = (w + u + iv)\exp(ib\xi)$, where $u,v$ are the real and imaginary parts of perturbation that can grow with a complex rate $\delta$ upon propagation. Standard linearization procedure leads to the following eigenvalue problem

$$\delta u = -\frac{1}{2}\frac{d^2 v}{d\eta^2} - nv + bv,$$
$$\delta v = \frac{1}{2}\frac{d^2 u}{d\eta^2} + nu + \Delta n w - bu,$$
(2)

where $\Delta n = 2\int_0^L w(\lambda)u(\lambda)G(\eta,\lambda)d\lambda$ is the refractive index perturbation. The linear stability analysis shows that fundamental surface waves are always stable. Importantly, higher-order surface waves might be stable too. Thus, dipole waves exhibit complex, multiple stability domains (Fig. 4(a)). The widest stability domain is found to be the one that is adjacent to the upper cutoff. The width of stability domains gradually decreases when $b$ shifts deeper into the existence domain, so that when $b$ becomes smaller than a critical value the localized wave becomes unstable. The width of stability domains increases with increasing $n_b$ (see Fig. 4(c) that shows first three widest stability domains for the dipole surface waves on the plane $(n_b, n_b - b)$).

A similar picture was encountered for higher-order waves, but it should be stressed that the number of stability domains and their widths rapidly decrease with increase of the order of surface waves. Thus, triple-mode surface wave possesses only two stability domains at $n_b = 20$, in contrast to dipole wave having five stability domains for the same boundary refractive index value (Fig. 4(b)).

The results of linear stability analysis were confirmed by the direct simulations of propagation of surface waves perturbed by broadband noise with variance $\sigma_{\text{noise}}^2$. Stable representatives of surface wave families keep their internal structures over indefinitely long distances (Fig. 5), while unstable higher-order surface waves typically transform into fundamental waves via progressively increasing oscillations of their poles. Both fundamental and higher-order surface waves may be excited with single or several Gaussian beams with properly adjusted amplitudes and widths, launched at a point slightly shifted into the thermal medium (as in Fig. 1).

Summarizing, we predicted the existence of stable fundamental and higher-order surface waves at the interface of thermal media with defocusing nonlinearity whose edges are maintained at different temperatures. The physics behind the process is the competition between light repulsion by the interface and by the defocusing nonlinearity.

*Visiting from the Universidad de las Americas, Puebla, Mexico.

# References with titles

# References without titles

# Figure captions

Figure 1. Dynamics of propagation of a Gaussian beam $q|_{\xi=0} = a\exp[-(\eta-\eta_c)^2/\eta_w^2]$ launched in the center of thermal medium corresponding to $\eta_c = 20 = L/2$ (a) and at a distance $\eta_c = 6$ (b) from the left interface of thermal medium for $a = 1$, $\eta_w = 2$, and $n_b = 80$.

Figure 2. Profiles of (a) fundamental surface waves at $b = 11$ (1) and 17 (2), (b) dipole surface waves at $b = 13.3$ (1) and 16.6 (2), and (c) triple-mode surface waves at $b = 13.9$ (1) and 17 (2). In panels (a)-(c) refractive index $n_b = 20$. (d) The nonlinear corrections to refractive index for fundamental surface wave with $b = 3$ (a), 1.2 (2), and $-0.5$ (3) at $n_b = 4$.

Figure 3. (a) Energy flow versus propagation constant for fundamental surface wave. (b) Cutoff versus refractive index at the left boundary of thermal medium for fundamental surface wave (curve 0) and dipole surface wave (curve 1).

Figure 4. Real part of perturbation growth rate for dipole (a) and triple-mode (b) surface waves at $n_b = 20$. (c) Domains of stability (shaded) and instability (white) for dipole surface waves.

Figure 5. Stable propagation of fundamental surface wave corresponding to $b = 11$ (a) and dipole surface wave corresponding to $b = 13.3$ (b) in the presence of broadband input noise with variance $\sigma_{\text{noise}}^2 = 0.01$. In all cases $n_b = 20$.

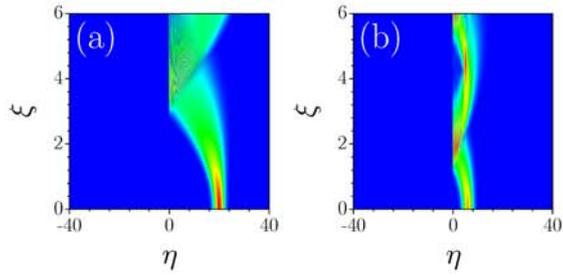

Figure 1. Dynamics of propagation of Gaussian beam $q|_{\xi=0} = a \exp[-(\eta - \eta_c)^2 / \eta_w^2]$ launched in the center of thermal medium corresponding to $\eta_c = 20 = L/2$ (a) and at a distance $\eta_c = 6$ (b) from the left interface of thermal medium for $a = 1$, $\eta_w = 2$, and $n_b = 80$.

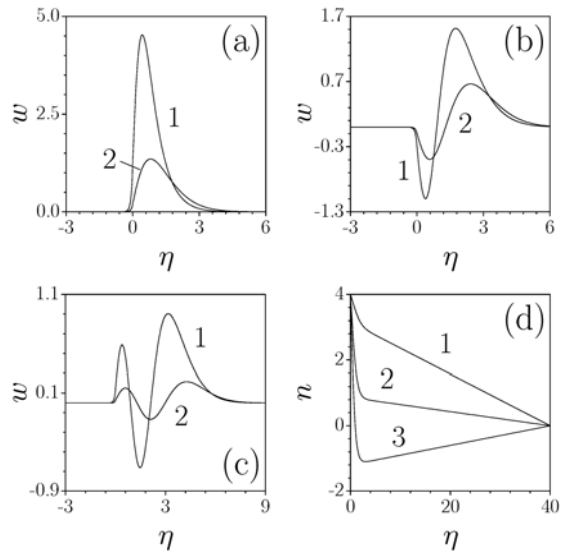

Figure 2.  Profiles of (a) fundamental surface waves at $b = 11$ (1) and 17 (2), (b) dipole surface waves at $b = 13.3$ (1) and 16.6 (2), and (c) triple-mode surface waves at $b = 13.9$ (1) and 17 (2). In panels (a)-(c) refractive index $n_b = 20$. (d) The nonlinear corrections to refractive index for fundamental surface wave with $b = 3$ (a), 1.2 (2), and $-0.5$ (3) at $n_b = 4$.

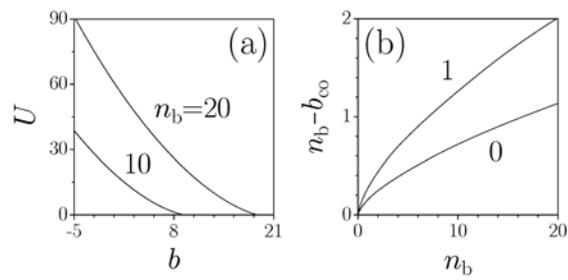

Figure 3. (a) Energy flow versus propagation constant for fundamental surface wave. (b) Cutoff versus refractive index at the left boundary of thermal medium for fundamental surface wave (curve 0) and dipole surface wave (curve 1).

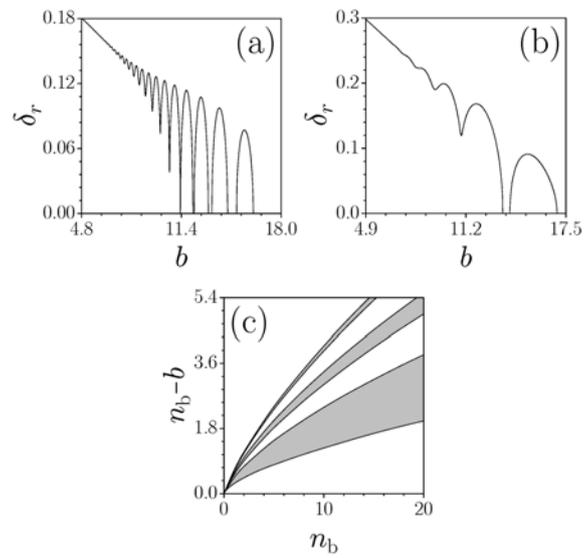

Figure 4. Real part of perturbation growth rate for dipole (a) and triple-mode (b) surface waves at $n_b = 20$. (c) Domains of stability (shaded) and instability (white) for dipole surface waves.

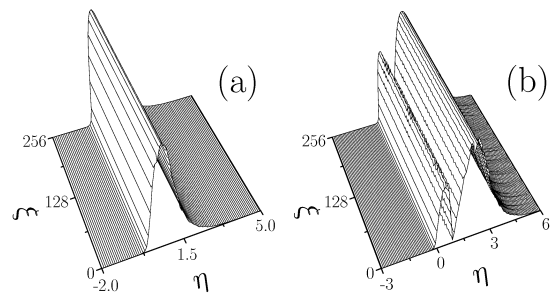

Figure 5. Stable propagation of fundamental surface wave corresponding to $b = 11$ (a) and dipole surface wave corresponding to $b = 13.3$ (b) in the presence of broadband input noise with variance $\sigma_{\text{noise}}^2 = 0.01$. In all cases $n_{\text{b}} = 20$.